# Quantum Communication for Military Applications

Niels M. P. Neumann[1], Maran P. P. van Heesch[1], Patrick de Graaf[2]

1: Cyber Security & Robustness,
The Netherlands Organisation for Applied Scientific Research (TNO), The Hague, The Netherlands
2: Information and Sensor Systems,
The Netherlands Organisation for Applied Scientific Research (TNO), The Hague, The Netherlands

Corresponding addresses: {niels.neumann, maran.vanheesch, patrick.degraaf}@tno.nl

*Abstract*—**Communication is vital in everyday life and critical for current and future military operations. However, conventional communication as we know it also has its limitations. Quantum communication allows some of these challenges to be overcome and, thereby, new application areas open up, also in the military domain. In this work quantum communication for military purposes is considered. Different applications are presented and the state-of-the-art of the technology is given. Also, quantum communication use cases specific for military applications are described.**

## I. Introduction

Communication is a vital part of everyday life, whether it is used for navigation, transaction systems or observing outer space. The same is true for military operations, where communication is a critical factor for achieving information dominance. The entire sensor-to-shooter-loop depends on robust and secure communications.

Conventional communication can be understood in the framework of classical physics. Information is for instance transmitted using optics via radio frequencies, free-space or via cables.

Conventional communication systems have reached a high level of maturity. The overhead is low as are the costs. However, conventional communication also presents some limitations in terms of for instance availability, bandwidth, direction (or lack of it) and security: Communication channels are typically secured by relying on computational hardness assumptions; Global Navigation Satellite Systems (GNSS) satellites like GPS or Galileo can only be synchronized up to a practically unavoidable amount of error; and, for similar reasons, ensembles of high resolution imaging telescopes cannot be too far apart limiting their capabilities. By basing communication on quantum mechanical principles, some of these limitations can be overcome and new possibilities arise.

Quantum mechanics stems from the beginning of the twentieth century and studies physical systems at small scales, e.g., molecules and atoms. These microscopic systems behave very differently from macroscopic systems. Current technological developments actively manipulate quantum mechanical systems and rely on their properties. These technologies arise in the interrelated fields of quantum computing, quantum communication and quantum sensing. The focus of this work is on quantum communication only.

Theoretically, applications of quantum communication have been long known, with as a prime example the BB84 quantum key distribution protocol [1]. Yet, exploiting the full potential of quantum communication is only just starting.

Two key principles of quantum mechanics, highly relevant for quantum communication, are superposition and entanglement. First, superposition is a property of states of quantum mechanical systems, like a particle. It means that although a system is definitely in one state, it can also be considered as being in several states. This is a non-classical concept since a classical system is always in a definite state specifying, e.g., the position of a particle. In contrast, quantum mechanical systems generally are complex linear combinations of these definite states. A quantum measurement is a probabilistic operation where a single output is obtained from the respective possible results with certain corresponding probabilities. The corresponding possible output states are the states present in the superposition. When a specific result for the measured observable is found, the original state collapses into the corresponding state.

Second, two quantum mechanical systems are entangled if one system cannot be fully described without referring to the other. Such a system is described by one, non-separable state. Manipulating one of the subsystems will have an immediate consequence on the other subsystems, independent on how far these subsystems are apart. Moreover, the results of measurements of observables in an entangled system will have correlations beyond what is possible in classical mechanics.

This quantum entanglement, together with the properties of superposition, are used in various applications of quantum communication networks. Examples of applications are: quantum key distribution [1], clock synchronisation [2], quantum interferometry [3], and position-based cryptography [4]. A full network of nodes that can exchange quantum information and entanglement, allowing for the implementation of all these applications, is also referred to as a quantum internet [5].

This paper focusses on the applications of quantum communication networks for military applications. Secure communication, accurate navigation and precise observation (sensing) are of the highest importance for military applications and quantum communication offers improvements to these capabilities.

In Section II an overview of different applications is given as well as an explanation of how quantum communication can be beneficial. In Section III the state-of-the-art of these technologies is given. Section IV presents use cases of quantum communication that might be applicable for military applications in the future. Conclusions and recommendations are given in Section V.

## II. Technology application overview

There are different applications of quantum communication. In this section different applications and potential benefits of quantum communication are introduced.

First, quantum key distribution is explained in Section II.A. Second, improved interferometry and distributed quantum computing are considered in Section II.B and Section II.C, respectively. Clock synchronisation as application is explained in Section II.D and position-based cryptography in Section II.E. Finally, the quantum internet is discussed in Section II.F.

### A. *Quantum key distribution*

One of the first applications of quantum communication is quantum key distribution (QKD), which enables two parties, Alice and Bob, to obtain a shared key between the two of them. There are two main approaches to quantum key distribution, *prepare-and-measure* and *entanglement-based*. Early adaptions of the former approach are already available on the market, though developments are still on going towards more robust solutions.

One of the key benefits of quantum key distribution is that for both prepare-and-measure and entanglement-based protocols, adversaries can be detected with high probability.

#### 1) *Prepare-and-measure QKD*

In prepare-and-measure QKD protocols, Alice prepares a, possibly superposed, quantum state, which she then sends to Bob. He measures the quantum state from which they can extract a key by using some classical post-processing techniques via authenticated conventional communication.

The first and most famous prepare-and-measure QKD protocol is the BB84-protocol, which stems from 1984 and is named after its inventors Bennett and Brassard [1]. In this protocol, Alice encodes bits of information in quantum states, using for instance photons. She sends these to Bob, who then measures the photons. Based on the choices of Alice and the measurement outcomes of Bob, they can extract a secure key.

The BB84-protocol provides information theoretic security, meaning that even an adversary with unlimited computational power cannot break the shared secret key. This in contrast to most conventional cryptography, which provide computational security.

Practical implementations can deviate from the theoretical protocol [6], which makes the use of the BB84-protocol in practice unsecure. Some of these practicalities can however be overcome, for instance by randomly varying the source intensity, as is done in the decoy-state protocol [7]. This still renders a secure protocol [8].

The performance of the decoy-state BB84-protocol is expressed by the maximum amount of key material one can extract per time frame. The performance is limited by the quality of the used hardware, the quantum communication channel and the used post-processing methods. The performance of the BB84-protocol can be determined and even optimised by carefully choosing model parameters [9].

Other prepare-and-measure quantum key distribution protocols work similarly [10]. A state is prepared by one party and subsequently measured by another party.

Typically, for prepare-and-measure QKD, fibre optic cables are used to send the encoded photons from Alice to Bob. However, it is also possible to send the photons via free-space, i.e. through the air. Even more interesting is the application where the decoy-state BB84-protocol is executed between a ground-station and a geo-stationary satellite [9]. These satellites follow the direction of the Earth's rotation with an orbital period equal to the Earth's rotational period.

For the security of a key, not only the key generation protocol is important, also the used devices are. Using compromised devices might render the results of a protocol insecure. In [9] a protocol is given that achieves independence from the used measurement devices with a prepare-and-measure QKD protocol.

*2) Entanglement-based QKD*

Instead of using single quantum states for quantum key distribution protocols, it is also possible to construct a QKD protocol using entanglement. The main idea of these protocols is that Alice and Bob share an entangled quantum pair which they use to extract key-material.

The work by Ekert is recognised to be one of the first QKD protocols using entanglement [12]. Here, different Bell-pairs, a specific quantum state, are shared between Alice and Bob. Both measure their part of the shared states randomly in one of three predetermined ways. Similar to the BB84-protocol, Alice and Bob now share how they measured, but not the measurement results. Based on this shared information they can extract a secure key.

Another protocol using entanglement is the BBM92-protocol [13], which can be seen as a generalisation of the BB84-protocol. Again, an entangled pair is distributed between Alice and Bob and both randomly choose one of two ways to measure their part of the entangled pair. If the entanglement generator in the BBM92-protocol is located at Alice and she directly measures her photon, this effectively results in the BB84-protocol.

There is a prepare-and-measure protocol that achieves independence from the used measurement devices. Using the entanglement-based BBM92-protocol, independence from the source can be achieved. In [14] a protocol is given to deal with imperfect devices. In theory, the protocol by Ekert [12] can achieve full device independency, however, this protocol is not yet possible in practice. Full device independency can be a desirable functionality, as one in this case does not have to trust the manufacturer of the hardware.

## B. *Improved interferometry*

Quantum communication can also be beneficial for various sensing applications. In astronomy, the power of the used telescopes depends on the effective angular resolution used. With a higher resolution, more details can be observed. However, the maximum achievable angular resolution is fundamentally limited by the wavelength $\lambda$ of the incoming light and the diameter $D$ of the aperture of the telescope:

$$\Delta\theta \simeq 1.22 \left(\frac{\lambda}{D}\right) \text{rad}$$

Note that a higher angular resolution means a lower value for $\Delta\theta$. As users cannot change the wavelength of the incoming light, the only way to achieve a higher angular resolution is by increasing the diameter of the aperture.

Instead of relying on one big telescope, an array of (smaller sized) telescopes can be used. By combining the light of all telescopes in a way that the phase information is preserved, interferometry can be performed. This combination is effectively equivalent to a telescope of larger diameter, while the total used surface is the sum of the surfaces of each individual telescope.

This is already applied. Classical receiver techniques are used to combine the incoming light from different telescopes at a central place and to create an interference pattern. This in turn gives a result with a higher angular resolution.

The main limitation of this approach is that the incoming light needs to be physically brought together at the central location without disturbing the phase. This is typically done using fibre optic links and the longer these links, the larger the impact of unforeseen phase shifts introduced in the path.

By using high-frequency recorders and highly synchronised clocks, interferometry is possible without the telescopes being physically connected. However, this poses requirements on the used hardware.

Instead, it is also possible to use quantum communication. By using shared entangled pairs, which are shared between telescopes using quantum communication, and letting the incoming light interact with the photons, interactions happen instantaneously and there is no need for highly synchronised clocks and high-frequency recorders. As the telescopes are in different positions, the incoming light has a relative phase between the two telescopes. This relative phase can be measured from the quantum state with high certainty, which in turn allows one to determine the source location precisely.

### C. Distributed quantum computation

Apart from quantum communication, quantum technology also offers new capabilities regarding computation using so-called quantum computers. Using quantum communication different quantum computers can be linked together [15][16]. This way, data from various parties can be used, without revealing the data to the others. Furthermore, computations requiring more resources than a single quantum device can offer, can be executed by distributing the computation over the available quantum computers. This could for instance be relevant for breaking conventional asymmetric cryptography using Shor's algorithms [17] and multiple early quantum computers.

### D. Clock synchronisation

Highly synchronised clocks are not just important for interferometry. Synchronised clocks are also used for syncing senders and receivers in various networks or for positioning and navigation purposes, where the signals of different satellites are combined to determine a location precisely.

When synchronising clocks between satellites, often the Poincaré-Einstein Synchronisation algorithm is used [18]. This algorithm uses light pulses to classically transmit timing information. However, this has an intrinsic inaccuracy due to random fluctuations in the refractive index of the atmosphere.

Another algorithm to synchronise clocks is Eddington's slow clock transport [19], where two clocks are synchronised in the same location and then transported to remote locations adiabatically slow. This however requires the exchange of a complex system, such as a clock. Due to time dilatation, the clocks can also not move at significantly different velocities as long as they will still be used.

Other clock synchronisation algorithms exist, however are all very similar to the two algorithms presented above. They therefore also all suffer from similar inaccuracies.

Quantum communication allows for perfect clock synchronisation, without errors introduced by transporting clocks or classically transmitting information.

The first quantum clock synchronisation algorithm was proposed by Jozsa et al. [2] in 2000. Even though the initial protocol raised some concerns such as requiring synchronised clocks to start with, these have been overcome.

The quantum clock synchronisation protocol still has some subtleties to overcome and there is still no consensus if all classical challenges are truly solved in practice.

### E. Position-based cryptography

Often, when data is encrypted, it can only be decrypted by using the key, a biometric property, or by performing a sufficiently strong brute force attack, where the latter is not guaranteed to succeed. However, there are also situations where information should be available not because one has the key, but because one is at a specific location. For instance, for special operations, where mission specific information should only be conditionally available: for a certain party, at a certain time and place. Key distribution should only be possible if the party is at that specific location. Cryptography where a key can only be shared between two parties at specific locations is called position-based cryptography.

Classically this form of cryptography is not possible if adversaries are allowed to cooperate [20]. Instead, when using quantum communication, position-based cryptography is possible, as was proven in [21]. One point of attention is that position-based cryptography is not possible at all if a group of adversaries is allowed to share an arbitrarily large entangled quantum state, as was also proven in [21].

*F. Quantum internet*

A quantum internet is a network of connected quantum computers and/or sensors. Different nodes are connected over the quantum network using shared entanglement. This shared entanglement has however its limitations. Generating entanglement over longer distances is increasingly hard and there is a maximum distance over which this shared entanglement can be generated sufficiently well. Therefore, quantum repeaters are used to achieve entanglement of sufficient quality over longer distances. The developments related to applications of quantum communication is closely related to the development of the quantum internet.

## III. State-of-the-art technology

A differentiation between sending single quantum states, such as individual photons, and quantum states consisting of two entangled photons is made for the description of the state-of-the-art technology of quantum communication technology. As stated in the previous section, various channels can be used for optical quantum communication; fibre, free-space (over ground or to and from space), and underwater. Even though the latter is a special case of free-space communication, it is listed as a separate channel, due to the potential application scenarios. The maturity level of these technologies is indicated in TABLE I.

TABLE I. Quantum communication technology level of maturity.

|  | Fibre | Free-space | Underwater |
|---|---|---|---|
| Single quantum states | High | Medium | Low |
| Entanglement | Medium | Low | - |

*1) State-of-the-art single photon transmission*

The quantum technology used for the single photon quantum states has as sole application QKD. Fibre-based prepare-and-measure QKD is already commercially available. This does not mean that this comes without limitations: the current distance limit that can be reached commercially is approximately 70 km with an expected key rate of 10kb/s. In [22] an experiment on prepare-and-measure QKD is performed where the maximal distance over which secure key-material is obtained is 421 km. At a distance of 405 km, a bit rate of 6.5 bits/seconds is achieved.

Prepare-and-measure QKD using free-space communication is currently in the research phase. No commercial applications are available yet, although they are expected on the short term. There are various ground-breaking experiments conducted to demonstrate free-space prepare-and-measure QKD. An overview of these ground-breaking results is shown in TABLE II.

Experiments to demonstrate the use of underwater QKD have been conducted, specifically with the application towards submarines in mind. The authors of [28] executed the BB84-protocol over a 2.37 m water channel and achieved a low quantum bit error rate of less than 3.5%, with different attenuation coefficients. A simulation conducted in [29] showed the potential of underwater QKD. In this work an oceanic channel is studied for various distances (10-100 m) and depths ranging from 100 m to 200 m. The distances largely depend on the possibilities and limitations of underwater laser communications.

TABLE II. Free-space decoy-state BB84 experiment results

| Experiment | Year | Distance | Key Rate | Ref. |
|---|---|---|---|---|
| Horizontal key exchange between static points | 2007 | 144 km | 42 b/s | [23] |
| Airplane in flight to ground-based receiver | 2013 | 20 km | 7.9 b/s | [24] |
| Static transmitter to moving receiver | 2015 | 650 m | 40 b/s | [25] |
| Ground-based transmitter to airplane in flight | 2017 | 3-10 km | 868 kb[1] | [26] |
| Satellite to static ground-based receiver | 2017 | 1200 km (ground distance) | 1.1 kbit/s | [27] |

*2) State-of-the-art entanglement-based transmission*

The first loophole-free Bell test experiment in 2015 proved that quantum entanglement exists without additional assumptions or loopholes that results up until then had [30]. Since then, quantum technology that is used for entanglement distribution has matured. As explained in Section II, quantum entanglement is used for more applications than just QKD.

Considering entanglement transmission through optical fibre, the maximal distance covered is 300 km by using time-bin entangled photon pairs [31]. In [32] results are presented on the observation of entanglement between matter (a trapped ion) and light (a photon) over 50 km of optical fibre. There are currently no experiments published in which entanglement is distributed underwater. One publication considers a 96 km long submarine optical fibre [33], which is an optical telecommunications fibre located underwater.

In the free-space prepare-and-measure QKD experiment reported in [27], entanglement has also been distributed over the established quantum communication link. Furthermore, Chinese researchers managed to show entanglement distribution from a satellite to the ground over a total ground distance of 1200 km [34].

Though distances are nearing interesting levels for military applications, the key rate is not always as impressive. It is however clear that progress is being made and live application in the foreseeable future is not imaginary.

*3) Supporting quantum hardware*

Research is needed to overcome the distance limitations and other challenges that limit the performance of quantum communication. These challenges include short coherence times of quantum information. Currently, research is done to improve the existing hardware and researchers also work to develop quantum repeaters and quantum memories [35].

A vision on the development of the quantum internet – and the expected order of realisation of various quantum technologies – is given in [36]. Each stage indicates a level of technology maturity and examples of known applications. The speed in which the technology is developed depends on many factors, which makes it hard to predict the speed in which this will go. The indicated stages of [36] are also shown in TABLE III.

---

[1] Maximal achieved key length.

TABLE III.   Different stages of development of a quantum internet. Table taken from [36].

| Stage | Functionality | New applications |
|---|---|---|
| 1 | Trusted repeater | Quantum key distribution (no end-to-end security) |
| 2 | Prepare and measure | Quantum key distribution, secure identification |
| 3 | Entanglement generation | Device independent protocols |
| 4 | Quantum memory | Blind quantum computing, simple leader election and agreement protocols |
| 5 | Few qubit fault tolerant computers | Clock synchronization, distributed quantum computing |
| 6 | Fault tolerant quantum computing | Leader election, quantum computing algorithms |

### IV. Military use cases

In the previous sections different applications of quantum communication were introduced and the state-of-the-art of quantum technology was considered. In this section, different use cases relevant for defence are considered.

Four different military use cases will be considered. The first use case on secure communication is presented in Section IV.A. The second use case is based on synchronised clocks and is given in Section IV.B. In Section IV.C and Section IV.D, use cases for position-based cryptography and improved sensors are considered, respectively.

#### A. Secure communication

As mentioned earlier, future quantum computers will be able to break current widely used asymmetric cryptographic protocols using Shor's algorithm. This will require new, quantum-safe approaches for encryption. Likewise, symmetric keys will be targeted with Grover's algorithm [37], though doubling the key length will probably ensure a quantum-safe situation for symmetric cryptography.

This is very relevant in the military domain. Information dominance will be lost and missions will be endangered if communications are not reliable, robust and secure anymore. To achieve secure communication, usually symmetric encryption is used. This symmetric encryption in turn requires keys to both encode and decode data. It is however not always possible to distribute these keys between all participating parties before a mission starts, or to change keys very frequently during a mission. In order to still safely communicate, a shared secret key must be exchanged. Using quantum key distribution protocols, this shared key can be generated in a theoretically secure manner.

Consider the situation illustrated in Fig. 1, where multiple possible scenarios for quantum key distribution in a military setting are shown. Most of the shown communication links use free-space quantum key distribution, including (laser) communication between ships (Link 1), satellites (Link 2), underwater vessels (Link 3), planes and/or drones (Link 4), and ground-based stations (Link 5) and vehicles (Link 7). Furthermore, a fibre optic communication channel between two ground-based stations is shown (Link 6). Fibre optic communication also allows for quantum key distribution between land-based strategic assets, such as headquarters, command posts, naval stations, bases and airfields, which usually have a fixed or semi-static location.

Each of the shown links is a possible application of quantum key distribution between two parties. However, each of the shown situations also comes with its own specific challenges. For example, for free-space quantum key distribution a direct line-of-sight is a necessity.

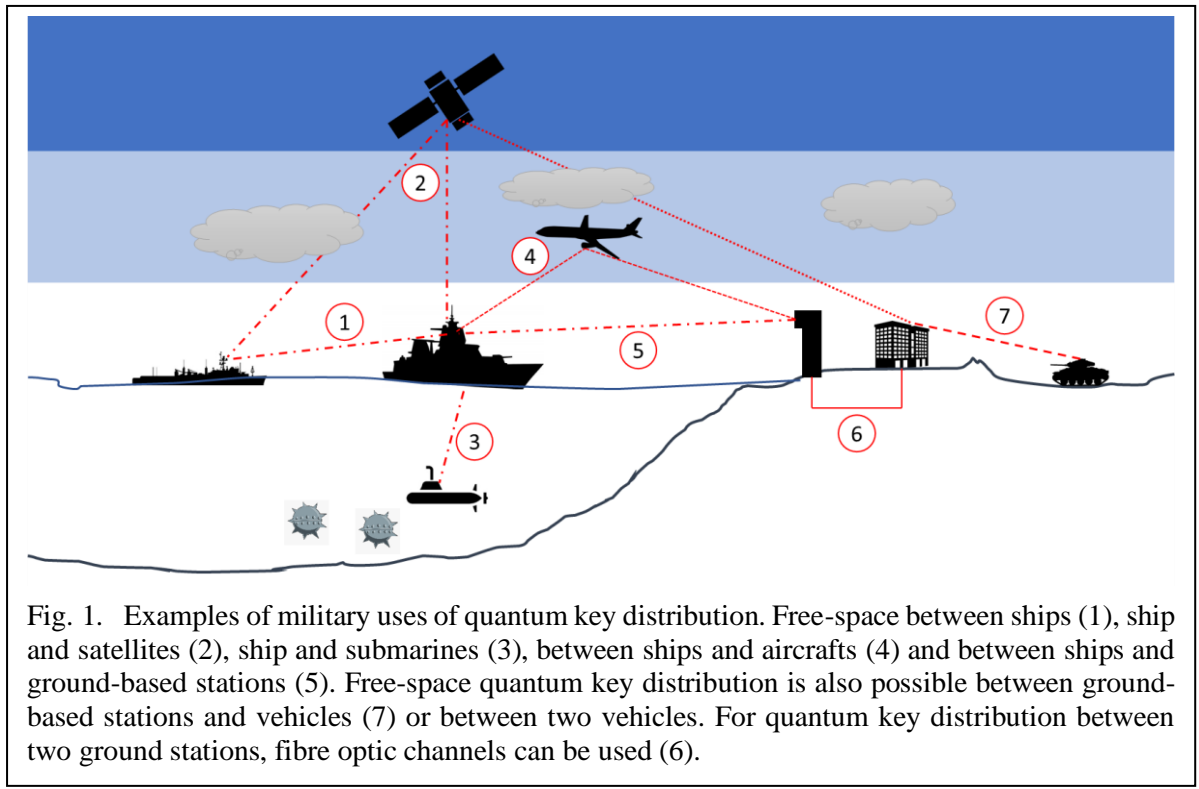

Fig. 1. Examples of military uses of quantum key distribution. Free-space between ships (1), ship and satellites (2), ship and submarines (3), between ships and aircrafts (4) and between ships and ground-based stations (5). Free-space quantum key distribution is also possible between ground-based stations and vehicles (7) or between two vehicles. For quantum key distribution between two ground stations, fibre optic channels can be used (6).

Different experiments have been performed for potential military applications of quantum key distribution, even though the experiments themselves showcase QKD without a particular user application in mind. The experiments listed in Section III and TABLE II. can be linked to the scenarios sketched in Fig. 1. Depending on whether the ships and land-based vehicles are moving or not, the following relations are shown: The results of [23] are applicable to Links 1, 3, 5, 6 and 7; The results of both [24] and [26] are applicable to Link 4; The results of [25] can be used for Links 1, 3, 5, 6 and 7; and, the results of [27] relate to Link 2.

Quantum key distribution is just one of the options for quantum-safe communication. Instead, also trusted couriers can be used and/or post-quantum cryptography. Post-quantum cryptography does not make use of quantum technologies, but instead is based on different hard mathematical problems that are expected to remain safe against quantum computers and hence burden adversaries with increased computational power requirements. When using post-quantum cryptography, the already available IT infrastructure, used by current encryption mechanisms, can be used, which eases the transition from one to the other encryption protocol. Note that both the trusted courier solution and post-quantum cryptography come with their own challenges and limitations.

### B. *Exact position determination*

Exact position determination is of the utmost importance when it comes to military missions, especially accurately determining the location of friendly units, guiding precision munition, or operating autonomous or unmanned vehicles (UxVs). This position can accurately be determined using GNSS. By perfectly synchronized clocks, the precision of GNSS can be made arbitrarily accurate.

Currently, most of the navigation and positioning determination is based on GNSS-satellites. This is in turn can however be spoofed, though very hard in practice, or access can be denied by jamming. For these scenarios it is not known if quantum communication offers solutions, however, the field of quantum sensing certainly does.

### C. *Position-based access to data*

The key idea of position-based cryptography is that data can only be decrypted by being on a specific location. As this application is new, it currently is not used in practice. It would also require a different approach to data access management than currently used, so not just role-based (RBAC), but conditionally- or risk-based access.

One potential application can be that data is only available if you are at the compound. This protects the data from malicious parties that have no access to the compound. Another use case would be the other way around: specific information should only become available at a certain place and a certain time during a sensitive mission.

### D. *Improved sensing capabilities*

For military application many different sensors are used, for instance to detect the use of certain transmitters. With larger detectors and telescopes, more accurate signals can be detected and noise can be suppressed more efficiently.

The technique currently used is a phased array technology, where detectors are distributed over different locations. Increasing the area over which the detectors are scattered, increases the detection capabilities and the detection precision. Furthermore, this scattering potentially allows for a better protection against failure or blockade of specific detectors in specific regions.

When scattering detectors over a larger area, the complexity of combining the incoming signals increases and the impact of noise will become too large for practical use. By using quantum communication, detectors over longer distances can coherently be combined.

## V. Conclusions

Quantum communication allows the improvement of various military applications and even creates some new opportunities. In this article a couple of potential use cases are highlighted. Other use cases can easily be envisioned as well, for instance making free-space communication less detectable by using quantum communication. This however completely depends on the characteristics of laser communication as transmission channel.

Although the implementation of quantum communications in today's operations is not yet possible, the field of quantum hardware development is very active and it is hard to predict how fast developments will go. It is clear however, that both allies and geopolitical competitors are investing substantially in quantum communication technology, with China's seven billion US dollar investment program as a notable example.

It is critical that military operations can benefit, as soon as possible, from quantum communication, and with that also from quantum sensing and quantum computing. This is in order to achieve and maintain information dominance under challenging conditions, facing (near) peer opponents utilising that kind of technology as well. By being involved in national and international quantum research collaborations, the ministries of defence can ensure that they are amongst the first to benefit from the new quantum technologies. That does not necessarily mean investing the basis of quantum technology itself. There is enough drive in the civil domain for that. The focus should be on future military applications. This way, developments can be steered in such a way that the most important technologies for military applications are indeed developed with some priority, such that these technologies are indeed fit for use in the military domain.

### ACKNOWLEDGEMENTS

This work was supported by the V1902 research program 'Tactical networks for mobile, dismounted and on foot operations' which is performed by TNO on behalf of The Netherlands Ministry of Defence.